\documentclass[useAMS,usenatbib]{mn2e}
\usepackage{ulem}
\usepackage[dvips]{graphicx}

\title[Fuzzy Dark Matter in Relativistic Stars]{
Fuzzy Dark Matter in Relativistic Stars}
\author[Zeinab Rezaei]{Zeinab Rezaei\thanks{E-mail:
	zrezaei@shirazu.ac.ir} \\
Department of Physics, School of Science, Shiraz
University, Shiraz 71454, Iran.\\
Biruni Observatory, School of Science, Shiraz
University, Shiraz 71454, Iran.}

\begin{document}

\date{Accepted XXX Received XXX }

\maketitle
	\begin{abstract}
Fuzzy dark matter (FDM), a practical alternative to cold dark matter, can exist in compact stars.
Here, applying the FDM equation of state (EoS) constrained by CMB and large-scale structure data,
we calculate the structure of relativistic stars in the presence of FDM.
For this aim, the EoS for the visible matter in neutron stars, quark stars, and hybrid stars from the observational data are employed.
A piecewise polytropic EoS constrained by the observational data of GW170817 and the data of six low-mass X-ray binaries with
thermonuclear burst or the symmetry energy of the nuclear interaction describes the neutron star matter.
For quark star matter, we apply the EoSs within the Bayesian statistical approach using the mass and
radius measurements of PSR J0030+0451 from NICER. Employing the two-fluid formalism, we study the structure of FDM admixed relativistic stars.
\end{abstract}


\section{Introduction}

Fuzzy dark matter (FDM) composed of ultralight bosonic particles with $m \sim 10^{-22} eV$ has been proposed to solve
different problems such as disagreement between cold dark matter (DM) predictions and small scale observations,
missing satellite problem and core-cusp problem in dwarf galaxies \citep{Khlopov,Hu,Hui,Burkert,Niemeyer}.
FDM as a Bose Einstein condensate with the quantum effects at scales in the order of kpc (the de Broglie wavelength of
particles) experiences quantum pressure as well as gravitational attraction.
Due to the balance among the quantum pressure and the gravity, a soliton core forms near the center of FDM halo
and the core structure can release the FDM particle properties \citep{Widrow}.
The behavior of FDM at large scales is not different from the cold DM, while the quantum nature of
FDM influences the structure formation at small scales \citep{Hu,Guth} and delays galaxy formation via
macroscopic quantum pressure \citep{Church}. The wavelike nature
of FDM results in the formation of granular structures in the FDM halo \citep{Kawai}.

Several studies have been considered to constrain the mass of FDM particles.
Galaxy luminosity function at high redshifts \citep{Schive,Menci},
Lyman alpha forests \citep{Armengaud,Irsic,Nori,Rogers},
CMB power spectrum \citep{Hlozek}, radius-dependent velocity dispersion \citep{Church}, abundance of Milky Way subhalos \citep{Nadler1,Nadler2},
tidal streams from globular clusters \citep{Dalal}, galactic ultra-faint
dwarf galaxies \citep{Hayashi}, observed displacements of star clusters and
active galactic nuclei from the centers of their host galaxies \citep{Chowdhury},
and the observations of high-redshift lensed galaxies from CLASH survey \citep{Kulkarni}
are some examples.
Ultralight axion DM is one of the candidate for FDM \citep{Svrcek,Dave}.
The forms of these axions have been predicted in string theory \citep{Cicoli}.
In some investigations, the detection of axion DM has been considered \citep{Abel}.

FDM can influence the astrophysical objects in different scales.
First galaxies are collected in a FDM cosmology and the primordial stars can form along dense DM filaments \citep{Mocz}.
The structure of self gravitating systems containing axions (axion stars) has been investigated and
the collision of axion stars with neutron stars (NSs) can release the energy of axions \citep{Barranco}.
There may be a large number of axion stars in galaxies and their collisions with each other and with other astrophysical
objects such as ordinary stars and NSs are possible \citep{Eby}.
The attractive self-interactions of DM axions result in nongravitational growth of density fluctuations and the formation of bound
objects can influence the axion density perturbations on length scales \citep{Arvanitaki}.
Cold DM axions may be converted into photons in the NS magnetosphere \citep{Huang,Foster,Battye}.
Axion DM can be detected via the narrow radio lines radiated by the NSs \citep{Huang,Hook,Safdi,Foster}.
Pulsar timing array experiment has been suggested to detect the FDM signals \citep{Khmelnitsky,Porayko4,Martino,Porayko8,Kato,Nomura}.
FDM affects the dynamics of binary systems \citep{Nacir,Armaleo}.
Variations of the orbital parameters of binary systems induced by the perturbations of FDM have been studied \citep{Blas}.

Recently, DM in different compact objects such as NSs and quark stars (QSs) has been one of the
interesting subjects in astrophysics.
NSs can constrain the asymmetric DM \citep{Garani1,Ivanytskyi}.
Low-mass NSs can be formed from the accretion-induced collapse of DM admixed white dwarfs \citep{Leung}.
Spectroscopy measurements of NSs have been employed to detect DM \citep{Camargo,Maity}.
NSs admixed with DM and the constraints on DM properties from the observation of GW170817 \citep{Quddus}
have been explored.
DM particles can be captured by NSs and this leads to the NSs thermalization \citep{Bell1,Acevedo,Keung,Bell2,Garani2,Bell3,Bell4,Anzuini,Kumar}.
DM interactions with muons \citep{Garani3}, DM Admixed NSs with
the DM-nucleons interactions via Higgs portal \citep{Bhat}, and self-interacting bosonic DM \citep{Rafiei}
have been considered.
DM affects the nuclear matter parameters and the equations of state (EoSs) of nucleonic-matter \citep{Das1} and
the curvatures of the NS \citep{Das2}.
By modeling a massive NS with DM particles, the secondary component of GW190814
has been constrained \citep{Das3}.
The possibility of the fact that GW190814 is a bosonic DM admixed compact star has been studied \citep{Lee}.
Mass radius relation and second Love number of stars containing ordinary matter and non-self annihilating fermionic DM
have been calculated \citep{Dengler}.
The transmutation of NSs admixed with DM and gravitational collapse in the star centers result in
the formation of black holes with masses $M\approx 1 M_{\odot}$ \citep{Garani4}.
Dynamical evolution of DM admixed NSs with fermionic DM has been investigated \citep{Gleason}.

Self-annihilating neutralino WIMP DM may accrete into NSs and compact objects with long-lived lumps of
strange quark matter form \citep{PerezGarcia}.
The regions of stability for compact stars containing massless quark matter and fermionic DM have been calculated \citep{Mukhopadhyay}.
The observation of strange QSs could set constraints on
the scattering cross sections of light quarks and non-interacting scalar DM \citep{Zheng}.
The structure of strange stars admixed with self-interacting bosonic DM has been considered \citep{Panotopoulos1,Panotopoulos2,Lopes}.
The observations of strange stars in GW170817 confirmed that these stars have a mirror DM core \citep{Yang}.
According to the above discussions, one can easily conclude that the FDM can have important effects on relativistic stars.
In this paper, we study the structure of NSs, QSs, and hybrid stars in the presence of FDM.

\vspace{-0.58cm}
\section{Fuzzy dark matter constrained by the observational data}

\begin{figure*}
\vspace*{1cm}       
\includegraphics[width=0.3\textwidth]{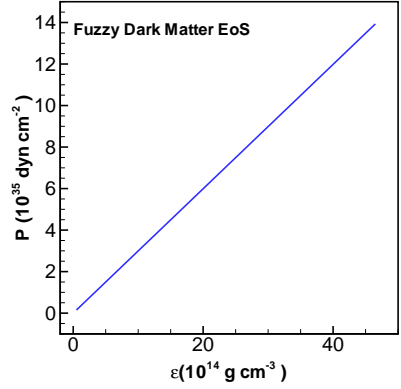}
\caption{Fuzzy dark matter EoS with the parameters $m=10^{-24} eV$ and $\lambda=10^{-100}$ from the observational data.}
\label{fig1}
\end{figure*}

In this study, we employ a constrained FDM model with a quartic self-interaction \citep{Cembranos}.
For this aim, a scalar field $\phi$, with Lagrangian
\begin{eqnarray}\label{}
   \mathcal{L}=\frac{1}{2}g^{\mu\nu}\partial_{\mu} \phi \partial_{\nu} \phi-V(\phi),
 \end{eqnarray}
 are considered. The potential has the form
\begin{eqnarray}\label{}
   V(\phi)=\frac{1}{2}m^{2}\phi^2+\frac{1}{4}\lambda\phi^4,
 \end{eqnarray}
in which $m$ denotes the mass term and $\lambda$ shows the strength of quartic self-interactions. Assuming a homogeneous and isotropic universe with a flat Robertson-Walker metric, anharmonic corrections to the mass term lead to the EoS for the scalar field with pressure $P$ and density $\rho$,
\begin{eqnarray}\label{fdmpress}
   w=\frac{P}{\rho},
 \end{eqnarray}
 with
 \begin{eqnarray}\label{}
   w=\frac{\frac{3\lambda}{8m^4}\rho}{1+\frac{9\lambda}{8m^4}\rho}.
 \end{eqnarray}
Applying CMB \citep{Ade} and large-scale structure (LSS) \citep{Parkinson} data, the parameters of this
model have been constrained \citep{Cembranos}. The constraint for the mass is $m\geq10^{-24} eV$ and
for allowed masses, the constraint on $\lambda$ is as follows,
\begin{eqnarray}\label{}
log_{10}\lambda<-91.86+4log_{10}(\frac{m}{10^{-22} eV}).
 \end{eqnarray}
Here, to describe FDM, we apply the values $m=10^{-24} eV$ for the mass and $\lambda=10^{-100}$ for the
self-interactions of FDM. In Figure. \ref{fig1}, we have presented the EoS of FDM
constraint with the observational data.



\vspace{-0.58cm}
\section{Two-fluid formalism for fuzzy dark matter admixed stars}

Starting with two-fluid formalism \citep{Sandin,Ciarcelluti}, we apply one static and spherically
symmetric spacetime described by the line element,
\begin{eqnarray}
      d\tau^2=e^{2\nu(r)}dt^2-e^{2\lambda(r)}dr^2-r^2(d\theta^2+sin^2\theta d\phi^2),
 \end{eqnarray}
and the energy momentum tensor of a perfect fluid,
\begin{eqnarray}
      T^{\mu \nu}=-p g^{\mu \nu}+(p+\varepsilon)u^{\mu}u^{\nu}.
 \end{eqnarray}
In the expression $T^{\mu \nu}$, $p$ and $\varepsilon$ are the total
pressure and total energy density, respectively, which are the
results of both visible ($V$) and dark ($D$) sectors,
  \begin{eqnarray}\label{press}
      p(r) = p_V(r) + p_D(r),
 \end{eqnarray}
   \begin{eqnarray}
      \varepsilon(r) =\varepsilon_V(r) + \varepsilon_D(r).
 \end{eqnarray}
In Eq. (\ref{press}), $p_V$ stands for the EoS of visible matter in compact stars, while $p_D$
presents the FDM EoS given by Eq. (\ref{fdmpress}).
Considering the above profiles, the Einstein field equations result in \citep{Sandin,Ciarcelluti}
\begin{eqnarray}\label{10}
      e^{-2\lambda(r)}=1-\frac{2M(r)}{r},
       \end{eqnarray}
\begin{eqnarray}
      \frac{d\nu}{dr}=\frac{M(r)+4\pi r^3 p(r)}{r[r-2M(r)]},
       \end{eqnarray}
\begin{eqnarray}
            \frac{dp_V}{dr}=-[p_V(r)+\varepsilon_V(r)] \frac{d\nu}{dr},
 \end{eqnarray}
\begin{eqnarray}\label{13}
            \frac{dp_D}{dr}=-[p_D(r)+\varepsilon_D(r)] \frac{d\nu}{dr}.
 \end{eqnarray}
Here, $M(r)=\int_0^r dr 4 \pi r^2 \varepsilon(r)$ denotes the total mass inside a sphere with radius $r$
and we specify the visible matter sphere and DM sphere with
the conditions $p_V(R_{V})=0$ and $p_D(R_{D})=0$, respectively. In this work, we assume that the densities of visible matter and dark matter are the same in the center of the star.

For the stars in the binaries, the tidal forces lead to induce the tidal deformabilities in the stars \citep{Hinderer0}. The traceless quadrupole moment tensor of the star $Q_{ij}$ is related to the tidal field tensor $E_{ij}$ by
\begin{eqnarray}
            Q_{ij}=-\frac{2}{3}k_2 R_{V}^5 E_{ij}=-\lambda E_{ij},
 \end{eqnarray}
in which $\lambda = \frac{2}{3}k_2 R_{V}^5$ denotes the tidal
deformability. Besides, the tidal Love number $k_2$ is as follows \citep{Hinderer8},
\begin{eqnarray}
            k_2&=&\frac{8 \beta^5}{5}(1-2\beta)^2[2-y_R+(y_R-1)2\beta]\nonumber \\&\times&[2\beta(6-3y_R+3\beta(5y_R-8))\nonumber \\&+&4\beta^3(13-11y_R+\beta(3y_R-2)+2\beta^2(1+y_R))\nonumber \\&+&3(1-2\beta)^2[2-y_R+2\beta(y_R-1)] ln(1-2\beta)]^{-1}.
 \end{eqnarray}
and $\beta=M/R$ presents the compactness of the star. Furthermore, solving the following differential
equation leads to the value of $y_R=y(r=R_V)$,
\begin{eqnarray}\label{tideq}
        r\frac{dy(r)}{dr}+y^2(r)+y(r)F(r)+r^2Q(r)=0.
 \end{eqnarray}
The functions $F(r)$ and $Q(r)$ are given by,
\begin{eqnarray}
        F(r)=[1-4\pi r^2(\varepsilon(r)-p(r))](1-\frac{2M(r)}{r})^{-1},
 \end{eqnarray}
and
\begin{eqnarray}
        r^2 Q(r)=4\pi r^2[5\varepsilon(r)+9p(r)+\frac{\varepsilon(r)+p(r)}{\partial p(r)/\partial \varepsilon(r)}]\nonumber \\ \times (1-\frac{2M(r)}{r})^{-1} -6 (1-\frac{2M(r)}{r})^{-1}\nonumber \\ - \frac{4 M^2(r)}{r^2}(1+\frac{4\pi r^3 p(r)}{M(r)})^2 (1-\frac{2M(r)}{r})^{-2},
 \end{eqnarray}
We solve Eq. (\ref{tideq}) along Eqs. (\ref{10})-(\ref{13}) with the initial condition $y(0) = 2$.
In addition, the dimensionless tidal deformability is defined by
\begin{eqnarray}
       \Lambda = \frac{2}{3}k_2\frac{R_V^5}{M^5}.
       \end{eqnarray}

In the case of quark star which is self bound, the discontinuity of the energy density at the surface of star should be considered. In the present study, we apply the boundary treatment on the stellar surface
to join the interior solution with the exterior one as in Refs. \citep{Damour,Postnikov,Zhou},
\begin{eqnarray}
        y_{R}^{ext}=  y_{R}^{int}-\frac{\varepsilon_s}{M/{4\pi R_V^3}},
         \end{eqnarray}
in which $\varepsilon_s$ is the energy density at the surface of star.

\vspace{-0.58cm}
\section{Fuzzy dark matter admixed neutron star}

In order to quantify the visible matter in NSs, we utilize the EoS
of dense NS matter in the form of a piecewise polytropic expansion
which is constrained by the observational data of GW170817 and the data of six low-mass X-ray binaries (LMXB) with thermonuclear burst or the symmetry energy of the nuclear interaction \citep{Jiang}.
The EoS with the expression $P=K\rho^{\Gamma}$ is parameterized with four pressure parameters
$\{\hat{p_1},\hat{p_2},\hat{p_3},\hat{p_4}\}$ at the corresponding densities of $\{1, 1.85, 3.7, 7.4\}\rho_{sat}$
in which the saturation density has the value $\rho_{sat}=2.7\times10^{14} gcm^{-3}$ \citep{Ozel}.
The joint analysis confirms that the constraint on $\hat{p_1}$ mainly is the result of nuclear constraints,
the constraint on $\hat{p_2}$ is predominantly determined by the gravitational wave data and the LMXB sources with thermonuclear bursts, the constraint on $\hat{p_3}$ heavily comes from the LMXB source data
and the current bounds of $M_{TOV}$, and the range of $\hat{p_4}$ is
narrowed down by LMXB sources with thermonuclear burst.
\begin{figure*}
\vspace*{1cm}       
\includegraphics[width=0.6\textwidth]{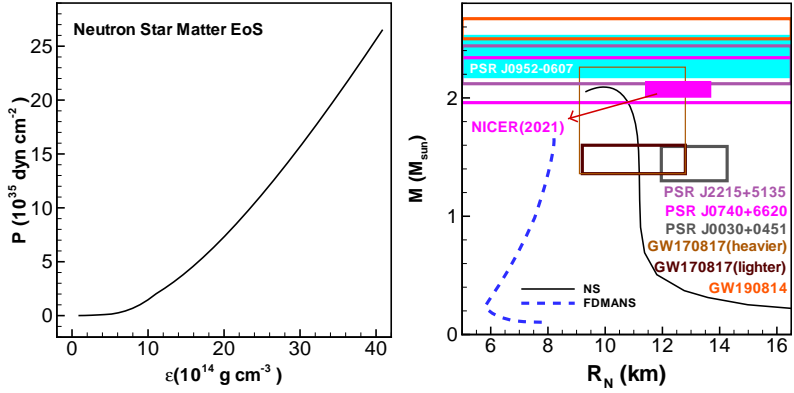}
\caption{Left: EoS of dense neutron star matter constrained by the observational data and Right:
Mass radius relation in the cases of neutron star (NS) and FDM admixed neutron star (FDMANS). Observational constraints on the NS radii and masses
from the pulsars and the gravitational wave data are also presented. These constraints are related to NICER observations for PSR J0952-0607 \citep{Romani}, PSR J2215+5135 \citep{Linares}, PSR J0740+6620 \citep{Cromartie,Fonseca,Miller21,Riley}, PSR J0030+0451 \citep{Miller}, and the merger events GW170817 \citep{Abbott7,Abbott8} and GW190814 \citep{Abbott}.
}
\label{fig2}
\end{figure*}

Piecewise polytropic EoS of NS matter and the mass radius relation for both NS and NS admixed with FDM are given in Figure. \ref{fig2}.
For the FDM admixed neutron star (FDMANS), we have considered the total mass versus the visible radius, i.e. the radius of sphere
containing NS matter.
FDM leads to stars with lower masses. The radius of FDMANSs is smaller than the
radius of NSs with the same mass. Therefore, FDM results in more compact stars.
For most FDMANSs, the larger stars are more massive, in contrary with NSs.
The interaction between FDM and NS matter leads to the self-bound FDMANSs a behavior different from the normal NSs which are gravitationally bound.
We have also shown the constraints on the mass radius relation obtained from the pulsars and the gravitational wave data with different colour bars.
NICER observations for PSR J0952-0607 \citep{Romani}, PSR J2215+5135 \citep{Linares}, PSR J0740+6620 \citep{Cromartie,Fonseca,Miller21,Riley}, PSR J0030+0451 \citep{Miller}, and the merger events GW170817 \citep{Abbott7,Abbott8} and GW190814 \citep{Abbott} give these constraints.
Both NSs and FDMANSs satisfy the constraints from the recent observational data.
The presented observations confirm that the maximum mass of FDMANSs is lower than the value $\sim 2.0 M_{\odot}$.
FDM leads to stars with lower maximum mass than all the observational data shown in this figure.

\begin{figure*}
\vspace*{1cm}       
\includegraphics[width=0.3\textwidth]{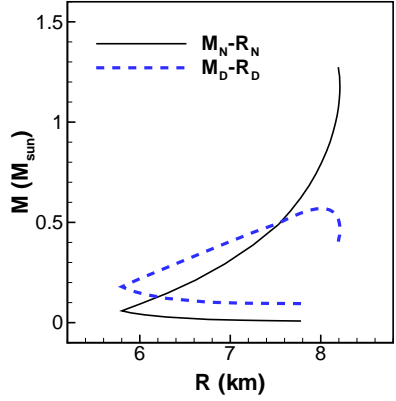}
\caption{Mass radius relation for two sectors of neutron star matter ($M_N-R_N$) and dark matter ($M_D-R_D$) in FDMANS.}
\label{fig3}
\end{figure*}

Figure. \ref{fig3} explains the behavior of visible and dark sectors in FDMANSs.
In very low mass stars, the mass of two sectors is not sensitive to the size of spheres.
However, for other FDMANSs, the mass of visible and dark spheres grows by increasing the radius.
The results confirm that for dark sphere, this behavior is not valid for all stars
and in large dark spheres, the mass decreases as the radius grows.
Figure. \ref{fig3} verifies that in smaller FDMANSs, the mass of dark sphere is higher than
the NS matter sphere. However, in larger FDMANSs, the mass of visible sector
is dominant. This opposite behavior for visible and dark sectors in FDMANS is due to the different EoSs of two sectors.

\begin{figure*}
\vspace*{1cm}       
\includegraphics[width=0.9\textwidth]{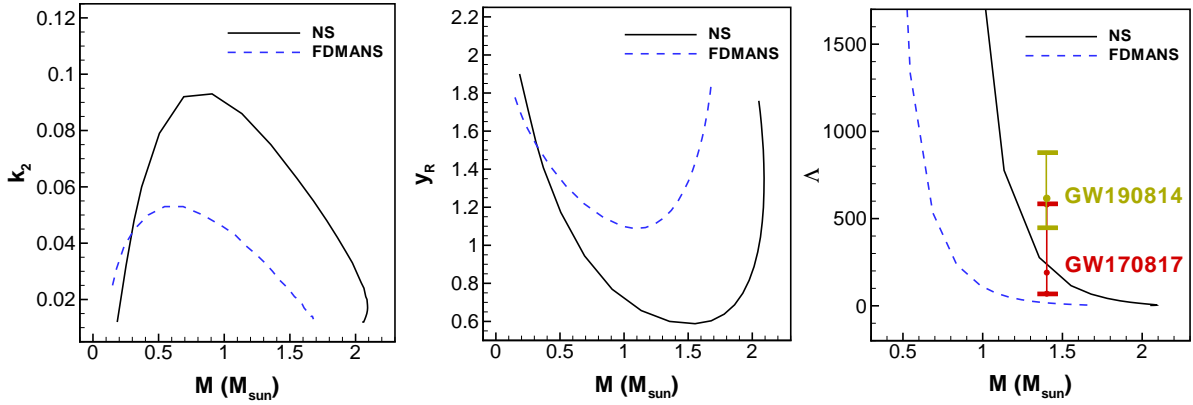}
\caption{Tidal Love number $k_2$, the value $y_R$, and the dimensionless tidal deformability $\Lambda$ versus the mass for NS and FDMANS.
The constraints from GW170817 and GW190814 data (LIGO and Virgo Collaborations) for neutron star of mass $M = 1.4 M_{\odot}$ are also given.
These upper limits on dimensionless tidal deformability are $\Lambda_{1.4}=190^{+390}_{-120}$ for GW170817 \citep{Abbott8} and $\Lambda_{1.4}=616^{+273}_{-158}$ for GW190814 \citep{Abbott}.
}
\label{figtidalns}
\end{figure*}

In Figure. \ref{figtidalns}, we have presented the tidal Love number $k_2$, the value $y_R$, and the dimensionless tidal deformability $\Lambda$ in the cases of NSs and FDMANSs. Except for the low mass stars, the tidal Love number decreases due to presence of the FDM in the stars. Besides, the star mass corresponding to the maximum value of the tidal Love number is lower when the FDM is considered in the stars. However, the value $y_R$ is higher in FDMANSs compared to NSs for most cases. Our calculations confirm that the dimensionless tidal deformability decreases with the star mass for both NSs and FDMANSs. The FDM leads to a considerable reduction of the dimensionless tidal deformability. This decrease is more significant in low mass stars. Moreover, we have shown the upper limits on dimensionless tidal deformability $\Lambda_{1.4}=190^{+390}_{-120}$ for GW170817 \citep{Abbott8}
and $\Lambda_{1.4}=616^{+273}_{-158}$ for GW190814 \citep{Abbott} obtained by LIGO and Virgo Collaborations.
In NSs, the dimensionless tidal deformability is in the range $70\leq\Lambda_{1.4}\leq580$ related to GW170817. This is while the parameter $\Lambda$ for NSs is lower than $\Lambda_{1.4}=616^{+273}_{-158}$ related to GW190814. Considering the FDMANSs, both upper limits from GW170817 and GW190814 are larger than the dimensionless tidal deformability.

\vspace{-0.58cm}
\section{Fuzzy dark matter admixed quark star}

In this work, we apply three EoSs of QSs within the Bayesian statistical approach using the mass and
radius measurements of PSR J0030+0451 from NICER \citep{Li}. These self-bound strange quark matter EoSs are based on the bag models in which the finite quark mass and superfluidity are also considered.
Our system describing the strange quark matter is a mixture of the massless u, d quarks and electrons, as well as s quarks of finite mass $m_s$ \citep{Haensel}.

In the first model, i.e. normal quark matter, the grand canonical potential per unit volume in the bag model
is expressed by,
\begin{eqnarray}
          \Omega_{Normal}=\sum_{i=u,d,s,e} \Omega_i^0+\frac{3(1-a_4)}{4\pi^2}\mu^4+B_{eff}.
 \end{eqnarray}
Here, $\Omega_i^0$ denotes the grand canonical potential for particle type $i$ as the ideal Fermi gas \citep{Farhi} and
$\mu=(\mu_u+\mu_d+\mu_s)/3$ presents the average quark chemical potential.
In addition, $B_{eff}$ determines the contributions from the quantum chromodynamics (QCD) vacuum,
and $a_4$ shows the perturbative QCD contribution from one-gluon exchange for gluon interaction.
Besides, the number density of each part of strange quark matter is related to the chemical potential $\mu_i(i=u,d,s,e)$ by,
\begin{eqnarray}
           n_i=-\frac{\partial\Omega}{\partial\mu_i}.
 \end{eqnarray}
The conditions for the quark matter at the equilibrium state are given by the weak interactions,
\begin{eqnarray}
          \mu_d=\mu_u+\mu_e,
 \end{eqnarray}
\begin{eqnarray}
          \mu_d=\mu_s.
 \end{eqnarray}
The condition of charge neutrality is also considered,
\begin{eqnarray}
         \frac{2}{3}n_u= \frac{1}{3}[n_d+n_s]+n_e.
 \end{eqnarray}
For normal quark matter, the pressure of quark matter at each value of $\mu$ is calculated by,
\begin{eqnarray}
        P_{Normal}=- \Omega_{Normal},
 \end{eqnarray}
and the energy density of quark matter is as follows,
\begin{eqnarray}
        \varepsilon_{Normal}= \Omega_{Normal}+\sum_{i=u,d,s,e} \mu_i n_i.
 \end{eqnarray}
In the two-parameter model Normal$(B_{eff}; a_4)$, the strange quark mass is fixed as
$m_s = 100\ MeV$ and the two parameters $(B_{eff}; a_4)$ are determined from the joint MSP J0740+6620 and PSR J0030+0451
analysis \citep{Li}.

The second model describing the superfluid quark matter is Color-Flavor Locked (CFL) in which an additional term related to the
pairing energy is added to the grand canonical potential \citep{Li},
\begin{eqnarray}
          \Omega_{CFL}= \Omega_{Normal}+\frac{3m_s^4-48\Delta^2\mu^2}{16\pi^2}.
 \end{eqnarray}
In the three-parameter model CFL$(B_{eff}; a_4; \Delta)$, as Normal model, the strange quark mass is $m_s = 100\ MeV$
and the three parameters $(B_{eff}; a_4; \Delta)$ are constrained by the observational data \citep{Li}.
The third model is four-parameter model CFLm$(B_{eff}; a_4; \Delta; m_s)$ in which the strange quark mass $m_s$ of the CFL superfluid quark matter is also constrained by the observational data \citep{Li}.

\begin{figure*}
\vspace*{1cm}       
\includegraphics[width=0.3\textwidth]{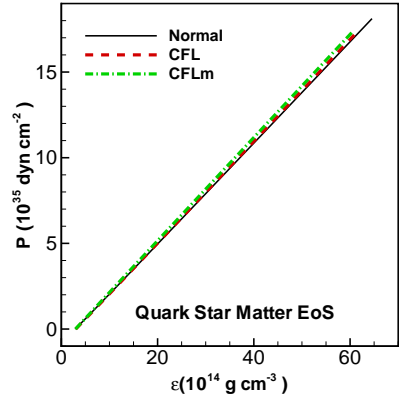}
\caption{Three EoSs of quark star matter based on the bag models constraint with the NICER data.}
\label{fig4}
\end{figure*}
\begin{figure*}
\vspace*{1cm}       
\includegraphics[width=0.9\textwidth]{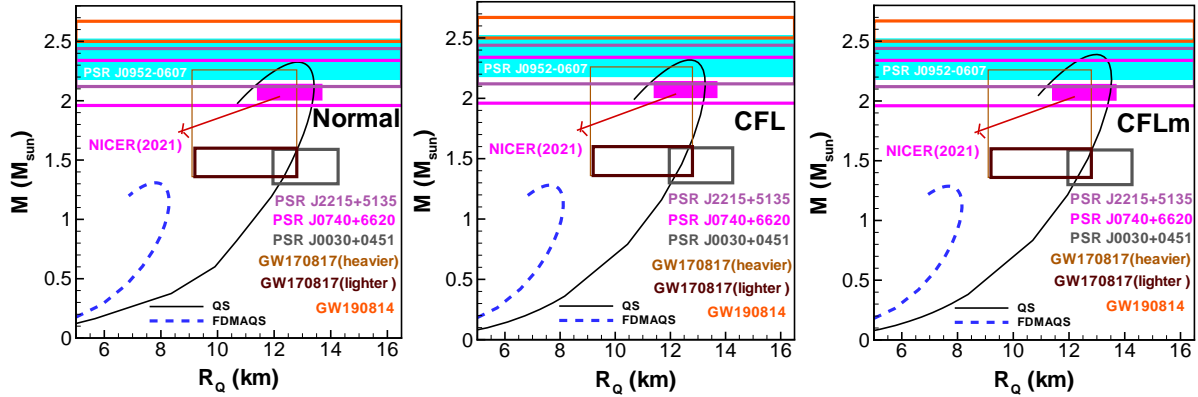}
\caption{Mass radius relation for quark star (QS) and FDM admixed quark star (FDMAQS) in three models for the EoS of quark star matter. Observational constraints are the same as Figure. \ref{fig2}.}
\label{fig5}
\end{figure*}

Figure. \ref{fig4} presents the three models for the EoS of strange star matter considered in this work.
In CFL and CFLm models, the EoS is stiffer than the EoS in Normal model.
CFLm model also leads to EoS which is stiffer than the EoS in CFL model. The mass radius relation for QSs and FDM admixed QSs (FDMAQSs) is given in Figure. \ref{fig5}.
In each model for quark matter, the maximum mass of FDMAQSs reaches the value lower than the one
for QSs. FDM affects the star so that the FDMAQSs are smaller than QSs with the
same mass. Therefore, the FDM leads to more compact stars, like the effect in NSs. This result is in agreement with the one obtained in \citep{Yang}.
QSs fulfill both the maximum mass and the mass radius constraints from the presented observational data. In addition, the maximum mass of FDMAQSs is lower than the value related to the maximum mass constraints.

\begin{figure*}
\vspace*{1cm}       
\includegraphics[width=0.9\textwidth]{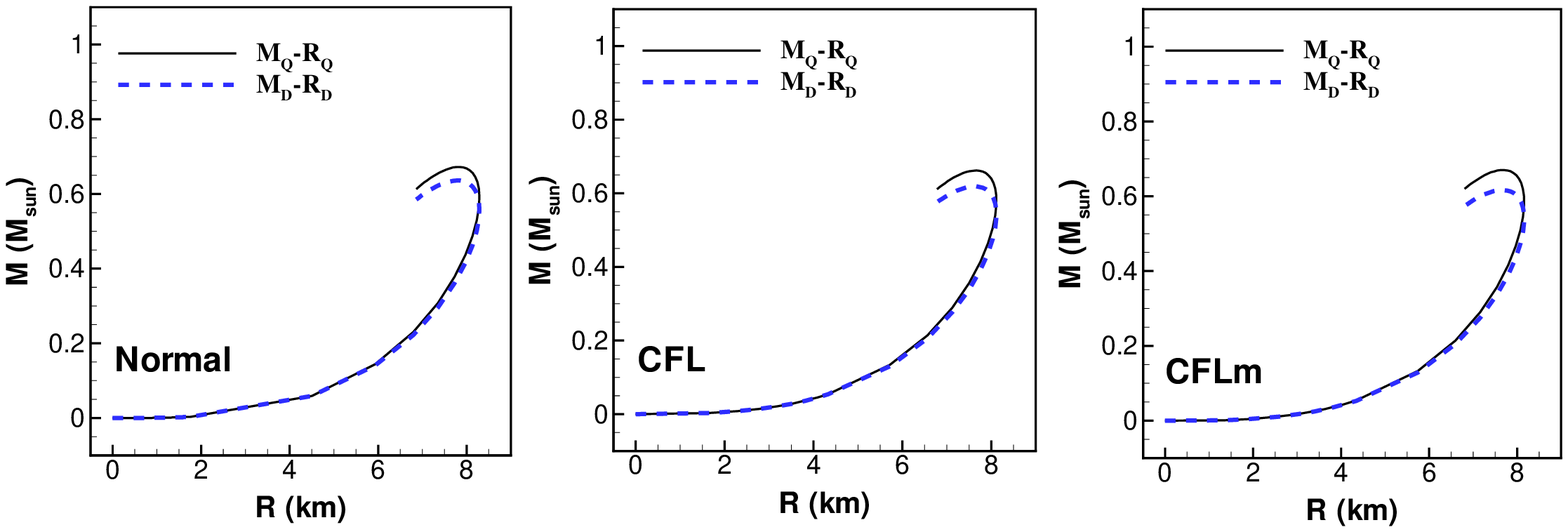}
\caption{Mass radius relation for two sectors of quark star matter ($M_Q-R_Q$) and dark matter ($M_D-R_D$) in FDMAQS.}
\label{fig6}
\end{figure*}

Figures. \ref{fig6} shows the mass radius relation for visible and dark sectors in FDMAQSs.
In three models, both visible and dark sectors represent a self-bound behavior like QS and FDMAQS.
For two spheres with smaller sizes, the mass of sphere is not sensitive to
the size. For most FDMAQSs, the contribution of two sectors in the mass of stars is the same. This is while in massive stars,
the mass of visible sphere is higher than the dark one.

\begin{figure*}
\vspace*{1cm}       
\includegraphics[width=0.9\textwidth]{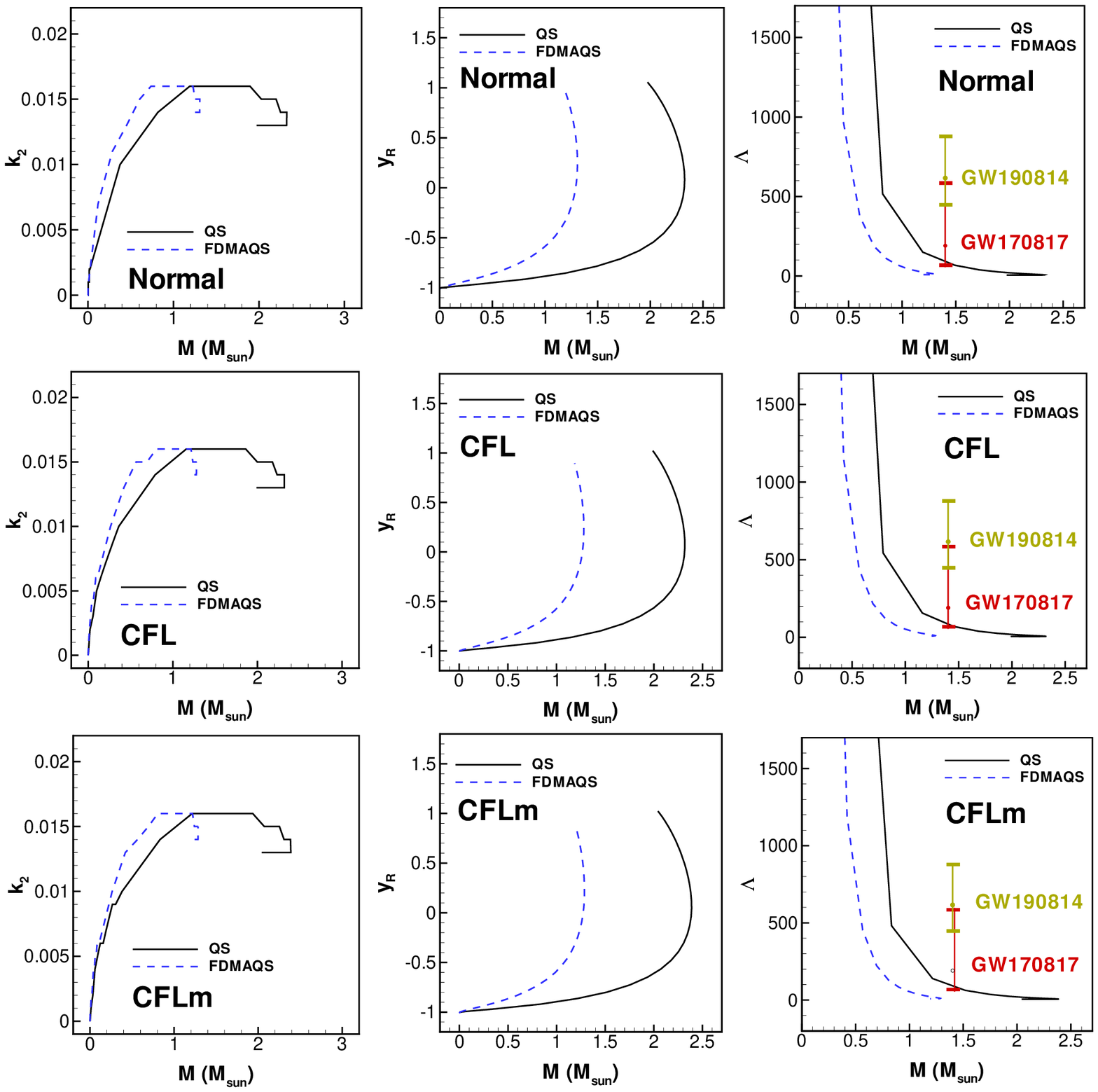}
\caption{Tidal Love number $k_2$, the value $y_R$, and the dimensionless tidal deformability $\Lambda$ versus the mass in the cases of
QS and FDMAQS in three models for the EoS of quark star matter. Observational constraints are the same as Figure. \ref{figtidalns}. }
\label{tidalqs}
\end{figure*}

The tidal Love number $k_2$, the value $y_R$, and the dimensionless tidal deformability $\Lambda$ for QSs and FDMAQSs are given in Figure. \ref{tidalqs}.
In most FDMAQSs, the tidal Love number takes higher values compared to the one in QSs with the same mass. This is while in massive FDMAQSs, FDM leads to the reduction of tidal Love number. Generally, FDMAQSs can experience larger values of the tidal Love number.
Figure. \ref{tidalqs} also indicates that for both QSs and FDMAQSs $y_R$ increases as the mass grows. $y_R$ for FDMAQSs is larger than for QSs. Our calculations confirm that FDM in QSs results in a considerable decrease of the dimensionless tidal deformability similar to the behavior in NSs. Besides, for both QSs and FDMAQSs, the dimensionless tidal deformability is lower than the upper limits from GW170817 and GW190814.

\section{Fuzzy dark matter admixed hybrid star}

For this study, we suppose that the hybrid star is composed of
a quark phase and a hadronic phase within a model like the one considered in \citep{Pereira1}. In our model, these two parts are split by a sharp phase-transition surface without a mixed phase and the density at the
phase-splitting surface can be discontinuous \citep{Pereira2}.

For the quark phase, we apply three EoSs, i.e. Normal, CFL, and CFLm models.
Furthermore, to describe the hadronic phase, the EoS of dense NS matter based on the observational data which considered in section 4 is applied. The density jump at the surface of quark-hadronic phase transition
is taken as a free parameter. By defining the parameter,
\begin{eqnarray}
        \eta\equiv\frac{\epsilon_q}{\epsilon_h}-1,
 \end{eqnarray}
in which $\epsilon_q$ shows the density at the top of the quark phase and $\epsilon_h$ denotes
the density at the bottom of the hadronic phase, we quantify the density jump.
According to $p_q = p_h$ at the quark-hadronic phase transition interface, $\epsilon_q$ or $\epsilon_h$ and the phase transition pressure are determined.
\begin{figure*}
\vspace*{1cm}       
\includegraphics[width=0.9\textwidth]{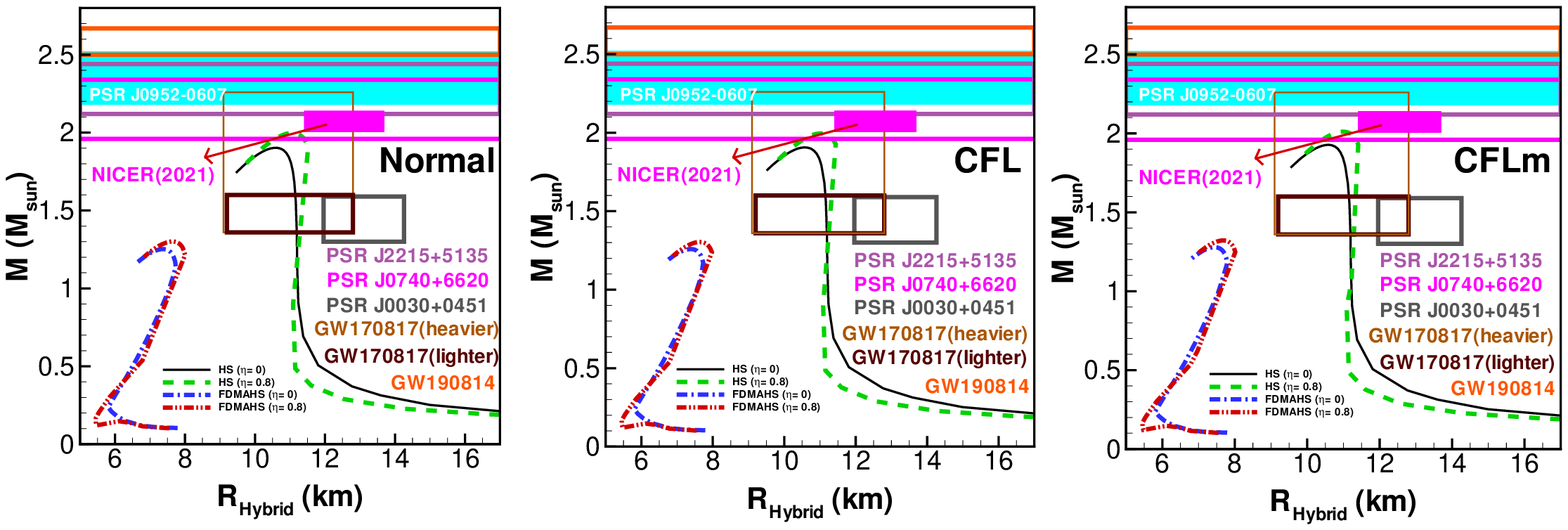}
\caption{Mass radius relation for hybrid star (HS) and FDM admixed hybrid star (FDMAHS) for two cases $\eta=0$ and $\eta=0.8$ in different models of quark matter EoS for the quark phase. Observational constraints are the same as Figure. \ref{fig2}}.
\label{fig8}
\end{figure*}

In Figure. \ref{fig8}, we have presented the mass radius relation for hybrid star and FDM admixed hybrid star (FDMAHS) in two cases $\eta=0$ and $\eta=0.8$. FDM affects the mass of hybrid stars
in a way that the maximum mass decreases. Similar to other compact objects, the FDMAHSs are smaller in size compared to hybrid stars. The FDM results in more
compact stars. Similar to QSs, HSs also satisfy both the maximum mass and the mass radius constraints. Moreover, our results verify that FDMAHSs fulfill the maximum mass constraint.

\begin{figure*}
\vspace*{1cm}       
\includegraphics[width=0.9\textwidth]{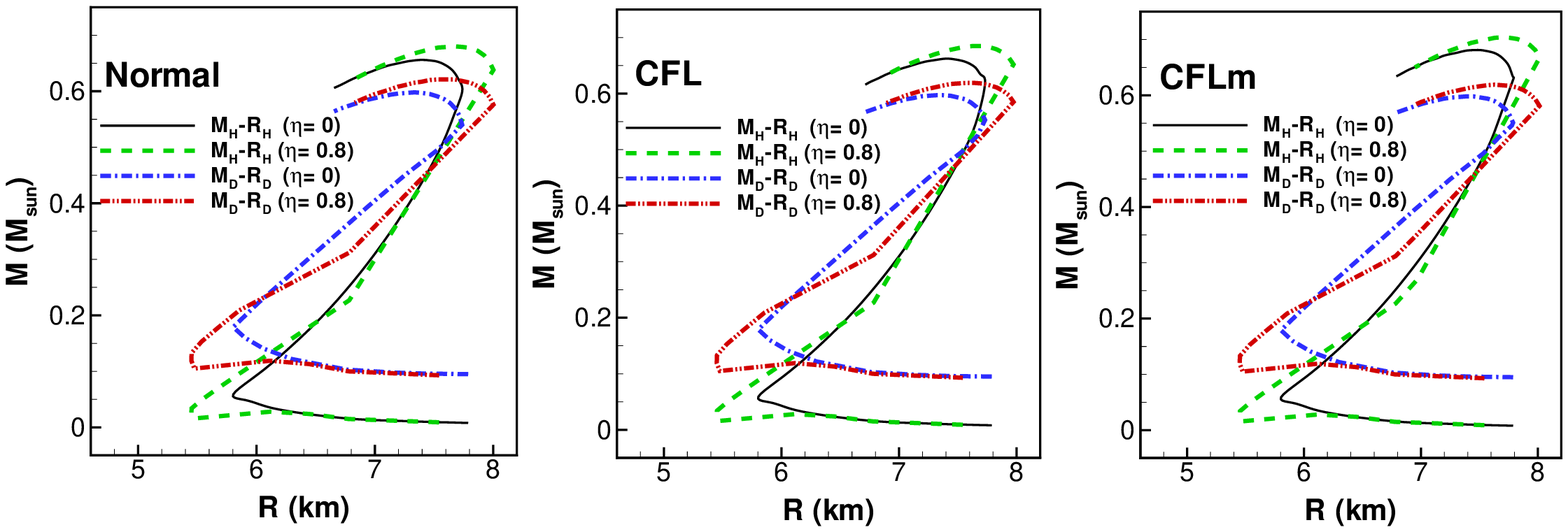}
\caption{Mass radius relation for two sectors of hadronic matter ($M_H-R_H$) and dark matter ($M_D-R_D$) in FDMAHS for two cases $\eta=0$ and $\eta=0.8$ in different models of quark matter EoS for the quark phase.}
\label{fig9}
\end{figure*}

Figure. \ref{fig9} gives the mass radius relations of visible and dark sectors in FDMAHSs.
The mass of sphere containing visible matter increases as the size grows.
In all models, the spheres are self-bound with different contributions of visible and dark matter in low and massive stars.
In discontinuous model, the range of the size of radius is higher than the one in continuous model. The low values of the mass of each visible and dark sectors show the contribution of these parts in the total mass of FDMAHSs. Besides, Figure. \ref{fig9} verifies that FDM results in the contraction of the FDMAHSs and smaller radius of stars.

\begin{figure*}
\vspace*{1cm}       
\includegraphics[width=0.9\textwidth]{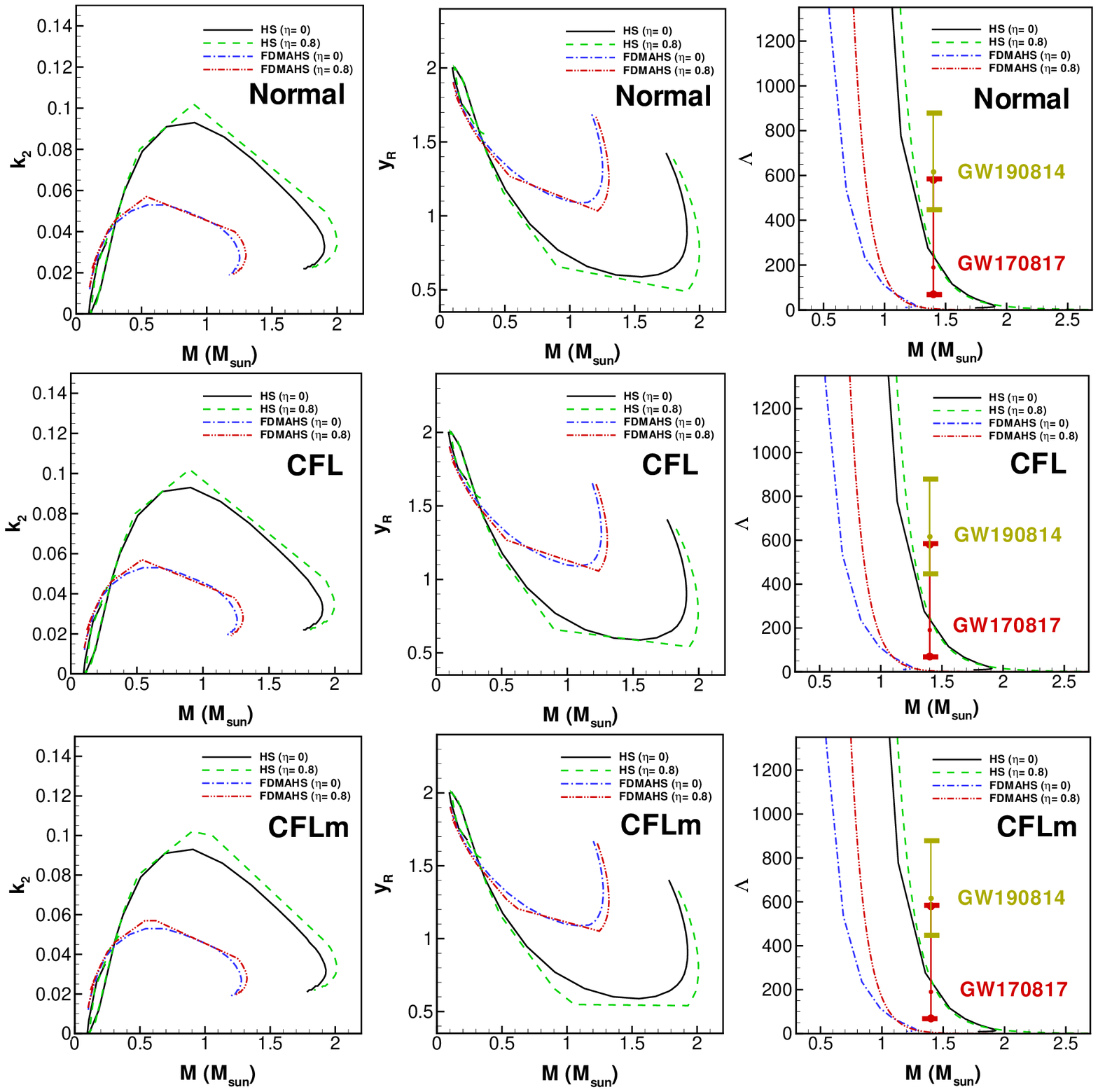}
\caption{Tidal Love number $k_2$, the value $y_R$, and the dimensionless tidal deformability $\Lambda$ versus the mass in HS and FDMAHS for two cases $\eta=0$ and $\eta=0.8$ in different models of quark matter EoS for the quark phase. Observational constraints are the same as Figure. \ref{figtidalns}.}
\label{tidalhs}
\end{figure*}

For HSs and FDMAHSs, we have shown the tidal Love number $k_2$, the value $y_R$, and the dimensionless tidal deformability $\Lambda$ in Figure. \ref{tidalhs}.
Except in low mass FDMAHSs, the tidal Love number of FDMAHSs is smaller than the one in HSs. Besides, considering HSs, the tidal Love number is larger in
discontinuous model. However, in FDMAHSs, the tidal Love number is almost the same in continuous and discontinuous models.
In addition, considering FDMAHSs, the value $y_R$ is smaller than the one in HSs. The discontinuous model gives lower values for $y_R$. Our calculations confirm that FDM considerably reduces the dimensionless tidal deformability of FDMAHSs like the one in FDMANSs and FDMAQSs. The dimensionless tidal deformability is higher in the discontinuous model compared to the continuous one. This enhancement is more significant in the low mass stars. Our calculations verify that for both HSs and FDMAHSs, the dimensionless tidal deformability is lower than the upper limits from GW170817 and GW190814.

\vspace{-0.58cm}
\section{Summary and Conclusions}
In the relativistic two-fluid formalism, we have explored the effects of fuzzy dark matter (FDM) on the
compact stars. The equations of state for FDM as well as the visible matter in stars which have been used
are based on the observational data.
Our results verify that in FDM admixed neutron stars, FDM leads to neutron stars with lower masses.
Moreover, FDM makes more compact neutron stars.
In FDM admixed neutron stars, the mass of visible and dark spheres grows as the radius increases.
Besides, the mass of visible and dark spheres depends on the size of the stars.
FDM admixed quark stars are smaller than quark stars without FDM with the
same mass and therefore they are more compact, like the phenomena in neutron stars.
FDM admixed hybrid stars are also more compact in comparison with hybrid stars with no FDM.
Furthermore, FDM in compact stars leads to a significant change in the dimensionless tidal deformability of stars.

\section*{Acknowledgements}
The author wishes to thank the Shiraz University Research Council.

\section*{DATA AVAILABILITY}
All data are given either in this paper or in the references.

\label{lastpage}
\vspace{-0.58cm}
\bibliographystyle{mn2e}

\end{document}